\documentstyle[preprint,aps,epsf]{revtex}
\begin{document}
\draft
\tighten
\title{Pions in Nuclei and Manifestations of Supersymmetry in
     Neutrinoless Double Beta Decay.}
\author{
Amand Faessler$^1$,  Sergey Kovalenko$^{1,2}$,
Fedor \v Simkovic$^{3}$}
\address{
1. Institute f\"ur Theoretische Physik der Universit\"at T\"ubingen,\\
   Auf der Morgenstelle 14, D-72076 T\"ubingen, Germany\\
2. Joint Institute for Nuclear Research, 141980 Dubna, Russia \\
3. Department of Nuclear Physics, Comenius University, \\
   Mlynsk\'a dolina F1, 84215 Bratislava, Slovakia}
\date{\today}
\maketitle
\begin{abstract}
 We examine the pion realization of the short ranged
supersymmetric (SUSY) mechanism of neutrinoless double beta decay
($0\nu\beta\beta$-decay). It originates from the $R$-parity violating
quark-lepton interactions of the SUSY extensions of
the standard model of the electroweak interactions.
We argue that pions are dominant SUSY mediators in $0\nu\beta\beta$-decay.
The corresponding nuclear matrix elements for potentially
$0\nu\beta\beta$-decaying isotopes are calculated within
the proton-neutron renormalized quasiparticle random phase
approximation (pn-RQRPA).
We define those isotopes which are most sensitive to the SUSY
signal and outlook the present experimental situation
with the $0\nu\beta\beta$-decay searches for the SUSY. Upper limits on
the $R$-parity violating 1st generation Yukawa coupling
$\lambda'_{111}$ are derived from various $0\nu\beta\beta$-experiments.
\end{abstract}
\pacs{12.60.Jv, 11.30.Er, 11.30.Fs, 23.40.Bw}

\section{Introduction}
The observation of neutrinoless nuclear double beta decay
\begin{eqnarray}
\label{process}
(A, Z)\longrightarrow (A,Z+2) + 2 e^-
\end{eqnarray}
($0\nu\beta\beta$) would undoubtedly indicate the presence of
the new physics beyond the standard model (SM) of electroweak interactions.
However, as yet there is no experimental evidence
for this lepton-number violating ($\Delta L = 2$)exotic process.
On the other hand non-observation of $0\nu\beta\beta$-decay at certain experimental
sensitivity allows one to set limits on some parameters of
the new physics.
An unprecedented accuracy and precision of the modern $0\nu\beta\beta$-decay
experiments allows one in certain cases to push these limits
far out of reach of the accelerator and the other non-accelerator
experiments.

A well known example is given by the upper limit on the light effective
Majorana neutrino mass $\langle m_{M}^{\nu} \rangle$.
From the $0\nu\beta\beta$-decay experiments \cite{hdmo97}
it was found $\langle m_{M}^{\nu} \rangle \le {\cal O}(1.1 $ eV$)$
\cite{si97}.
Recall that the Majorana neutrino mass term violates the lepton number
$\Delta L = 2$. This is exactly what is necessary for $0\nu\beta\beta$-decay
to proceed via the virtual neutrino exchange between the two neutrons.
In this case the $0\nu\beta\beta$-decay amplitude is proportional to
$\langle m_{M}^{\nu} \rangle$.

The Majorana neutrino exchange is not the only possible mechanism of
$0\nu\beta\beta$-decay.
The lepton-number violating quark-lepton interactions of the R-parity
non-conserving supersymmetric extensions of the SM
($R_p \hspace{-1em}/\;\:$ SUSY) can also induce  this process
\cite{Moh86}-\cite{FKSS97}.
$R$-parity is a discrete multiplicative  $Z_2$ symmetry
defined as $R_p=(-1)^{3B+L+2S}$, where $S, B$ and $L$
are the spin, the baryon and the lepton quantum number.
At the level of renormalizable operators R-parity can be
explicitly violated by the trilinear and the bilinear terms in
the superpotential and in the soft SUSY breaking sector.

The impact of the R-parity violation on the low energy phenomenology is
twofold. First, it leads the lepton number and lepton flavor violating
interactions directly from the trilinear terms.
Second, bilinear terms generate the non-zero vacuum expectation value
for the sneutrino fields $\langle\tilde\nu_i\rangle\neq 0$
and cause neutrino-neutralino as well as electron-chargino mixing.
The mixing brings in the new lepton number and lepton flavor violating
interactions in the physical mass eigenstate basis.

The implications of the trilinear and the bilinear terms for
the $0\nu\beta\beta$-decay were previously considered
in Refs. \cite{Moh86}-\cite{FKSS97-1}
and in Ref. \cite{FKSS98-1} respectively.
The $0\nu\beta\beta$-decay proved to be very sensitive probe of
the new interactions predicted in
the $R_p \hspace{-1em}/\;\:$ SUSY \cite{HKK96}-\cite{FKSS98-1}.

In this paper we return to the phenomenology of the trilinear terms
and perform a comprehensive analysis of their contribution
to the $0\nu\beta\beta$-decay, paying special attention to
the hadronization of the corresponding quark interactions and
to the nuclear structure calculations.  We will show that
for the case of the trilinear terms
the stage of hadronization plays especially important role
in derivation of the short-ranged $R_p \hspace{-1em}/\;\:$ SUSY mechanism of
$0\nu\beta\beta$-decay.

In our paper Ref. \cite{FKSS97} we had considered
the two-pion realization of the underlying $\Delta L = 2$
quark-level $0\nu\beta\beta$-transition $d d\rightarrow u u + 2 e^-$.
It was found that the corresponding contribution to $0\nu\beta\beta$-decay
absolutely dominates over the conventional two nucleon mode realization.
In this paper we generalize the previous treatment of the hadronization
of the $\Delta L = 2$ quark operators.

Searching for tiny effects of the physics beyond the SM in $0\nu\beta\beta$-decay
requires a reliable treatment of the nuclear structure as well.
In this paper we present the results of our calculations within
the proton-neutron renormalized Quasiparticle Random Phase Approximation
(pn-RQRPA) \cite{pn-RQRPA}, \cite{TU}.
We are listing the nuclear matrix elements for the
$R_p \hspace{-1em}/\;\:$ SUSY
mechanism of the $0\nu\beta\beta$-decay and their specific values
for the experimentally interesting isotopes.

        We introduce new characteristic of $0\nu\beta\beta$-decaying
        isotope: its sensitivity to the $R_p \hspace{-1em}/\;\:$ SUSY
        signal.
        This characteristic depends only on the corresponding
        nuclear matrix element and the kinematical phase-space factor.
        We calculate the SUSY sensitivities of all experimentally
        interesting isotopes and on this basis estimate prospects
        for SUSY searches in $0\nu\beta\beta$-experiments.
Applying our approach, we determine presently most successful
$0\nu\beta\beta$-experiment which establishes the most
stringent constraints on the R-parity violating Yukawa couplings.

The paper is organized as follows. In the next section we shortly outline
the minimal supersymmetric standard model with the explicit R-parity
violation ($R_p \hspace{-1em}/\;\:$ MSSM) and show the effective
$\Delta L = 2$
Lagrangian which describes in this model
the quark-level $0\nu\beta\beta$-transition. In section 3 we derive
the corresponding effective Lagrangian at the hadronic level in terms
of the meson and the nucleon fields. Section 4 is devoted to derivation of
the nuclear $0\nu\beta\beta$-transition operators in non-relativistic
impulse approximation and to calculation of their matrix elements in
the pn-RQRPA approach. Section 5 deals with the constraints on
$R_p \hspace{-1em}/\;\:$ Yukawa couplings from various $0\nu\beta\beta$-experiments.

\section{$R_p \hspace{-1em}/\;\:$ SUSY induced $\Delta L = 2$ quark-lepton
interactions.}
Let us shortly outline the minimal supersymmetric
standard model with the explicit R-parity violation
($R_p \hspace{-1em}/\;\:$ MSSM).

For the minimal MSSM field contents the most general
gauge invariant form of the renormalizable superpotential is
\begin{eqnarray}
\label{sup_gen}
           W = W_{R_p} + W_{R_p \hspace{-0.8em}/\;\:},
\end{eqnarray}
where the $R_p$ conserving part has the standard MSSM form \cite{Haber}
\begin{eqnarray}
\label{R_p-cons}
         W_{R_p} = h_L H_1 L E^c + h_D H_1 Q D^c
                   + h_U H_2 Q U^c + \mu H_1 H_2.
\end{eqnarray}
Here $L$,  $Q$  stand for lepton and quark
doublet left-handed superfields while $ E^c, \  U^c,\   D^c$
for lepton and {\em up}, {\em down} quark singlet  superfields;
$H_1$ and $H_2$ are the Higgs doublet superfields
with  weak hypercharges $Y=-1, \ +1$, respectively.
Summation over the generations is implied.

The $R_p$ violating part of the superpotential (\ref{sup_gen})
can be written as ~\cite{Rp},
~\cite{hall}
\begin{eqnarray}
\label{W_rp}
W_{R_p \hspace{-0.8em}/\;\:} = \lambda_{ijk}L_i L_j E^c_k +
\lambda^{\prime}_{ijk}L_i Q_j D^c_k + \mu_j L_j H_2+
\lambda^{\prime\prime}_{ijk} U^c_i D^c_j D^c_k,
\end{eqnarray}
The coupling constants $\lambda$ ($\lambda^{\prime\prime}$) are
antisymmetric in the first (last) two indices.

The soft supersymmetry breaking part of the scalar potential
also contains the $R_p \hspace{-1em}/\;\:$-terms of the form:
\begin{eqnarray}
\label{V_rp}
V_{R_p \hspace{-0.8em}/\;\:}^{soft} = \Lambda_{ijk}\tilde L_i \tilde L_j
\tilde E_k^c +
\Lambda^{\prime}_{ijk}\tilde L_i \tilde Q_j \tilde D_k^c +
\Lambda^{\prime\prime}_{ijk}\tilde U_i^c \tilde D_j^c \tilde D_k^c
+&& \\ \nonumber
+ \tilde\mu_{2j}^2\tilde L_j H_2
+ \tilde\mu_{1j}^2\tilde L_j H^{\dagger}_1 + \mbox{H.c.}&&
\end{eqnarray}
In Eqs. (\ref{W_rp}), (\ref{V_rp}) the trilinear terms proportional to
$\lambda, \lambda^{\prime}, \Lambda, \Lambda^{\prime}$
and the bilinear terms
violate the lepton number while the trilinear terms proportional to
$\lambda^{\prime\prime}, \Lambda^{\prime\prime}$
violate baryon number conservation.

It is well known that the simultaneous presence of lepton and baryon
number violating terms in Eqs. (\ref{W_rp}), (\ref{V_rp})
(unless the couplings are very small)
leads to unsuppressed proton decay. Therefore, either only
the lepton or the baryon number violating couplings can be present.
Certain discrete symmetries such as the B-parity
\cite{hall}, \cite{B-parity} may originate
from the underlying high energy scale theory
and forbid dangerous combinations of these couplings.
Henceforth we simply set
$\lambda^{\prime\prime}= \Lambda^{\prime\prime}=0$.

The remaining R-parity conserving part of the soft SUSY breaking
sector includes the scalar field interactions
\begin{eqnarray}
\label{V_Soft}
V_{R_p}^{soft}= \sum_{i=scalars}^{}  m_{i}^{2} |\phi_i|^2 +
h_L A_L H_1 \tilde L \tilde{E}^c + h_D A_D H_1 \tilde Q \tilde{D}^c +&&
\\  \nonumber
+ h_U A_U H_2 \tilde Q \tilde{U}^c  +
\mu B H_1 H_2 + \mbox{ H.c.}&&
\end{eqnarray}
and the "soft"  gaugino mass terms
\begin{eqnarray}
\label{M_soft}
{\cal L}_{GM}\  = \ - \frac{1}{2}\left[M_{1}^{} \tilde B \tilde B +
 M_{2}^{} \tilde W^k \tilde W^k  + M_{3}^{} \tilde g^a \tilde g^a\right]
 -   \mbox{ H.c.}
\end{eqnarray}
As usual, $M_{3,2,1}$ denote the "soft" masses of the $SU(3)\times
SU(2)\times U(1)$ gauginos $\tilde g, \tilde W, \tilde B$ while $m_i$
stand for the masses of the scalar fields.
The gluino $\tilde g$ soft mass $M_3$
coincides in this framework with its physical mass denoted
hereafter as $m_{\tilde g} = M_3$.

As mentioned in the introduction, we concentrate on
the phenomenology of the trilinear R-parity violating terms
$L Q D^c$ in the superpotential (\ref{W_rp}) and
perform a comprehensive analysis of their contribution
to the $0\nu\beta\beta$-decay.

These terms lead to the following $\Delta L=1$ lepton-quark operators
\begin{eqnarray}
\label{lambda}
{\cal L}_{\lambda } &=&
\lambda _{ijk}^{\prime}[\tilde{\nu}_{_{iL}}\bar{d}_{_k} P_{L} d_{_j} +
\tilde{d}_{_{jL}}\bar{d}_{_k} P_{L} \nu _{_i} +
\tilde{d}_{_{kR}}\bar{d}_{_j} P_{R} \nu^c_{_i}-
\tilde{e}_{_{iL}}\bar{d}_{_k} P_{L} u_{_j}   \\ \nonumber
&-&\tilde{u}_{_{jL}}\bar{d}_{_k} P_{L} e_{_i}-
\tilde{d}_{_{kR}} \bar{u}_{_j} P_{R} e^c_{_i}]\ +\ \mbox{H.c.}
\end{eqnarray}
Here, as usual $P_{L,R} = (1 \mp \gamma_5)/2$.

Starting from this fundamental Lagrangian, one can derive
the low-energy effective Lagrangian \cite{HKK96}, which describes
the quark-level $0\nu\beta\beta$-transition $d d\rightarrow u u + 2e^-$.
Integrating out the heavy degrees of freedom, we come up with the formula:
\begin{eqnarray}
 {\cal L}_{qe}\ =
\label{ql}
\frac{G_F^2}{2 m_{_p}}~ \bar e (1 + \gamma_5) e^{\bf c}
\left[\eta^{PS}\ J_{PS}J_{PS}
- \frac{1}{4} \eta^T\   J_T^{\mu\nu} J_{T \mu\nu} \right].
\end{eqnarray}
These $\Delta L_e = 2$ lepton-number violating effective interactions
are induced by heavy SUSY particles exchange.
An example of the Feynman diagram contributing to
${\cal L}_{qe}$ is given in Fig. 1. A  complete
list of the relevant diagrams can be found in  \cite{HKK96}.
Compared to Ref. \cite{HKK96} we have properly taken into account
in the Lagrangian (\ref{ql}) the contribution of
the color octet currents.

The color-singlet hadronic currents in Eq.\ (\ref{ql}) are
\begin{eqnarray}
\label{currents}
J_{PS} = J_P + J_S, \ \ \
J_{P} =   \bar u^{\alpha} \gamma_5 d_{\alpha}, \ \ \ \
J_{S} =   \bar u^{\alpha} d_{\alpha}, \ \ \ \
J_T^{\mu \nu} = \bar u^{\alpha} \sigma^{\mu \nu}(1 + \gamma_5) d_{\alpha},
\end{eqnarray}
where $\alpha$ is the color index and
$\sigma^{\mu\nu} = (i/2)[\gamma^{\mu}, \gamma^{\nu}]$.

The effective lepton-number violating parameters $\eta^{PS}, \eta^{T}$
depend on the fundamental parameters of the $R_p \hspace{-1em}/\;\:$ MSSM
and can be written in the form
\begin{eqnarray}
\label{etaq}
\eta^{PS} &=&  \eta_{\chi\tilde e} + \eta_{\chi\tilde f} +
\eta_{\chi} + \eta_{\tilde g} + 7 \eta_{\tilde g}^{\prime}, \\
\label{eta}
\eta^{T} &=& \eta_{\chi} - \eta_{\chi\tilde f} + \eta_{\tilde g}
- \eta_{\tilde g}^{\prime},
\end{eqnarray}
where we denoted
\begin{eqnarray}
\label{eta_g}
\eta_{\tilde g} &=& \frac{\pi \alpha_s}{6}
\frac{\lambda^{'2}_{111}}{G_F^2 m_{\tilde d_R}^4} \frac{m_P}{m_{\tilde g}}\left[
1 + \left(\frac{m_{\tilde d_R}}{m_{\tilde u_L}}\right)^4\right]\\
\eta_{\chi} &=& \frac{ \pi \alpha_2}{2}
\frac{\lambda^{'2}_{111}}{G_F^2 m_{\tilde d_R}^4}
\sum_{i=1}^{4}\frac{m_P}{m_{\chi_i}}
\left[
\epsilon_{R i}^2(d) + \epsilon_{L i}^2(u)
\left(\frac{m_{\tilde d_R}}{m_{\tilde u_L}}\right)^4\right]\\
\eta_{\chi \tilde e} &=& 2 \pi \alpha_2
\frac{\lambda^{'2}_{111}}{G_F^2 m_{\tilde d_R}^4}
\left(\frac{m_{\tilde d_R}}{m_{\tilde e_L}}\right)^4
\sum_{i=1}^{4}\epsilon_{L i}^2(e)\frac{m_P}{m_{\chi_i}},\\
\eta'_{\tilde g} &=& \frac{\pi \alpha_s}{12}
\frac{\lambda^{'2}_{111}}{G_F^2 m_{\tilde d_R}^4}
\frac{m_P}{m_{\tilde g}}
\left(\frac{m_{\tilde d_R}}{m_{\tilde u_L}}\right)^2,\\
\label{eta_end}
\eta_{\chi \tilde f} &=& \frac{\pi \alpha_2 }{2}
\frac{\lambda^{'2}_{111}}{G_F^2 m_{\tilde d_R}^4}
\left(\frac{m_{\tilde d_R}}{m_{\tilde e_L}}\right)^2
\sum_{i=1}^{4}\frac{m_P}{m_{\chi_i}}
\left[\epsilon_{R i}(d) \epsilon_{L i}(e)  + \right.\\ \nonumber
&+& \left.\epsilon_{L i}(u) \epsilon_{R i}(d)
\left(\frac{m_{\tilde e_L}}{m_{\tilde u_L}}\right)^2
+ \epsilon_{L i}(u) \epsilon_{L i}(e)
\left(\frac{m_{\tilde d_R}}{m_{\tilde u_L}}\right)^2
\right].
\end{eqnarray}
In Eqs. (\ref{eta_g})-(\ref{eta_end}) we used the standard notations
$\alpha_2 = g_{2}^{2}/(4\pi)$ and $\alpha_s = g_{3}^{2}/(4\pi)$ for
the $SU(2)_L$ and $SU(3)_c$ gauge coupling constants. We also denoted
the gluino $\tilde g$ and the neutralinos $\chi_i$ masses
as $m_{\tilde g}$ and $m_{\chi_i}$  respectively.
The Majorana neutralinos $\chi_i$ are linear combinations of
the gaugino and higgsino fields
\begin{eqnarray}
\label{neut-1}
\chi_i = {\cal N}_{i1} \tilde{B} +  {\cal N}_{i2}  \tilde{W}^{3} +
{\cal N}_{i3} \tilde{H}_{1}^{0} + {\cal N}_{i4} \tilde{H}_{2}^{0}.
\end{eqnarray}
Here $\tilde{W}^{3}$ and $\tilde{B}$ are the neutral
$SU(2)_L$ and $U(1)$ gauginos while $\tilde{H}_{1}^{0}$,
$\tilde{H}_{2}^{0}$
are the higgsinos which are superpartners of the two
neutral Higgs boson fields $H_1^0$ and $H_2^0$
with weak hypercharges $Y=-1, \ +1$, respectively.

The matrix ${\cal N}_{ij}$, introduced in Eq. (\ref{neut-1}), rotates
the $4\times 4$ neutralino mass matrix $M_{\chi}$ to the diagonal
form $Diag[m_{\chi_i}]$. We define these matrix as in Ref. \cite{Haber}.

Neutralino couplings are defined as \cite{Haber}
\begin{eqnarray}
\label{coeff1}
\epsilon_{L i}(\psi) &=& - T_3(\psi) {\cal N}_{i2} +
                                \tan \theta_W \left(T_3(\psi)
                        -  Q(\psi)\right) {\cal N}_{i1},\\
\epsilon_{R i}(\psi) &=& Q(\psi) \tan \theta_W {\cal N}_{i1}.
\end{eqnarray}
Here $Q $ and $ \ T_3$ are the electric charge and the weak
isospin of the fields $\psi = u, d, e$.

\section{Hadronization of $R_p \hspace{-1em}/\;\:$ SUSY
quark-lepton interactions}

The next step deals with reformulation of the quark-lepton interactions
in Eq.\ (\ref{ql}) in terms of the effective hadron-lepton interactions.
This is necessary for the subsequent nuclear structure calculations.
%

There are the two possibilities of hadronization of the effective
Lagrangian ${\cal L}_{qe}$ in Eq.\ (\ref{ql}).
One can place the four quark fields in the two initial neutrons
and two final protons separately. This is the conventional 2N-mode of
$0\nu\beta\beta$-decay shown in Fig. 2(a).
Then $nn\rightarrow pp + 2e^-$ transition is directly induced by
the underlying quark subprocess
\begin{eqnarray}
\label{qtrans}
dd\rightarrow uu + 2e^-.
\end{eqnarray}
In this case the nucleon transition is mediated by the exchange of
a heavy supersymmetric particle like the gluino $\tilde g$
with the mass $m_{\tilde g}\ \mbox{${}^> \hspace*{-7pt} _\sim$} \ 100$GeV.
Therefore, the two decaying
neutrons are required to come up very closely to each other what
is suppressed by the nucleon-nucleon short range repulsion.

Another possibility is to incorporate quarks involved in the underlying
$R_p \hspace{-1em}/\;\:$ SUSY transition in Eq. (\ref{qtrans}) not
into nucleons but into two virtual pions \cite{FKSS97} or
into one pion as well as into one initial neutron and one final proton.
Now the $nn\rightarrow pp + 2e^-$ transition is mediated by the charged
pion-exchange between the decaying neutrons, as shown in Fig. 2(b,c).
This is what we call the one- and two-pion modes of $0\nu\beta\beta$-decay.
Since the interaction region extends to the distances  $\sim 1/m_{\pi}$
this mode is not suppressed by the short range nucleon-nucleon repulsion.
An additional enhancement of the $\pi$-modes comes from
the hadronization of the $R_p \hspace{-1em}/\;\:$ SUSY quark-lepton
vertex operator in Eq.\ (\ref{ql}) as discussed below.
In Ref. \cite{FKSS97} it was shown that the two-pion mode absolutely
dominates over the 2N-mode. In what follows, we are arguing that
it dominates over the one-pion mode as well.

The effective hadronic Lagrangian taking into account both
the nucleon (p, n) and $\pi$-meson degrees of freedoms in a nucleus can
be written as follows:
\begin{eqnarray}
\label{2pi}
&&  {\cal L}_{he} =
{\cal L}_{2N} + {\cal L}_{2\pi} + {\cal L}_{1\pi} + {\cal L}_s =
\frac{G_F^2}{2 m_{_p}}~
\bar p \Gamma^{(i)} n \cdot \bar p \Gamma^{(i)}  n \cdot
~\bar{e} (1 + \gamma_5) e^c -
\\  \nonumber
&&- \frac{G_F^2}{2 m_{_p}}~ m_{\pi}^2
\left[
 m_{\pi}^2 a_{2\pi} \left(\pi^-\right)^2
-  a_{1\pi} \bar p\ i \gamma_5 n\cdot \pi^-\right]\cdot
~\bar{e} (1 + \gamma_5) e^c +
g_{_s}~ \bar p\ i \gamma_5 n ~ \pi^+.
\end{eqnarray}
Here ${\cal L}_s$ stays for the standard pion-nucleon interaction with
the coupling $g_{_s} = 13.4\pm 1$ known from experiment.
The lepton-number violating terms
${\cal L}_{2N}$,  ${\cal L}_{1\pi}$, ${\cal L}_{2\pi}$
generate the conventional two-nucleon mode, the one and two
pion-exchange modes of the $0\nu\beta\beta$-decay respectively.
The corresponding diagrams are presented in Fig. 2.

The two-nucleon mode term ${\cal L}_{2N}$ with different operator
structures $\Gamma^{(i)}$ had been considered in
\cite{Verg87,HKK96}
within the $R_p \hspace{-1em}/\;\:$ MSSM. As was already mentioned
this term gives the sub-dominant contribution to $0\nu\beta\beta$-decay
in comparison with the pion terms \cite{FKSS97}.
Therefore, in this paper we concentrate on the effect of
the pion-exchange contribution generated by the terms
${\cal L}_{1\pi}$ and ${\cal L}_{2\pi}$ in Eq.\ (\ref{2pi}).

The basic parameters $a_{2\pi}$ and $a_{1\pi}$ of the Lagrangian
${\cal L}_{he}$ in Eq.\ (\ref{2pi})
can be approximately related to the parameters of the quark-lepton
Lagrangian ${\cal L}_{qe}$ in Eq. (\ref{ql}), using
the on-mass-shell "matching conditions" \cite{FKSS97}
\begin{eqnarray}
\label{q-had1}
<\pi^+,2e^-|{\cal L}_{qe}|\pi^-> &=&
<\pi^+,2e^-|{\cal L}_{2\pi}|\pi^->, \\
\label{q-had2}
<\pi^+,p,2e^-|{\cal L}_{qe}|n> &=&
<\pi^+,p,2e^-|{\cal L}_{1\pi}|n>.
\end{eqnarray}
In order to solve these equations we apply the widely used factorization
and vacuum dominance approximations \cite{okun}
for the matrix elements of the products of the two quark currents.
Then we obtain, taking properly into account
the combinatorial and color factors:
\begin{eqnarray}
\label{approx}
\nonumber
&&\langle\pi^+|J_{PS} J_{PS} |\pi^- \rangle \approx
\frac{5}{3} \langle\pi^+| J_P|0\rangle
\langle 0| J_P|\pi^- \rangle,\\
&&\langle\pi^+|J_T^{\mu\nu} J_{T \mu\nu} |\pi^-\rangle \approx
-4 \langle\pi^+| J_P|0\rangle
\langle 0| J_P|\pi^- \rangle,\\
&&\langle p|J_{PS} J_{PS} |n \pi^- \rangle \approx
\frac{5}{3} \langle p| J_P|n\rangle
\langle 0| J_P|\pi^- \rangle, \\
&&\langle p|J_T^{\mu\nu} J_{T \mu\nu} |n\pi^-\rangle \approx
-4 \langle p| J_P|n\rangle
\langle 0| J_P|\pi^- \rangle.
\end{eqnarray}
Here we applied the equalities
\begin{eqnarray}
\label{tens}
<0| J_{S}|\pi> = <0| J_{T}^{\mu\nu}|\pi(p_{\pi})> = 0.
\end{eqnarray}
The scalar matrix element vanishes due to the parity arguments,
the tensor one vanishes due to
$ J_{T}^{\mu\nu} = -  J_{T}^{\nu\mu}$ and impossibility of
constructing an antisymmetric object having  only one external
4-vector $p_{\pi}$.

We also use the following  relationships for the hadronic matrix elements:
\begin{eqnarray}
&&\langle 0|\bar u \gamma_5 d |\pi^-\rangle = i \sqrt{2} f_{\pi}
\frac{m_{\pi}^2}{m_u + m_d} \equiv i m_{\pi}^2 h_{\pi}, \\
&&\langle p| \bar u \gamma_5 d|n\rangle =
F_P \langle p|\bar p\gamma_5 n |n\rangle.
\label{pi-N-pi}
\end{eqnarray}
where $f_{\pi} = 0.668~ m_{\pi}$.
For the nucleon pseudoscalar constant $F_P$ we take its
bag model value $F_P \approx 4.41$  from Ref. \cite{adler}.

In this approximation we solve the "matching conditions"  in
Eqs.\ (\ref{q-had1})-(\ref{q-had2})
and determine the coefficients in Eq.\ (\ref{2pi})
\begin{eqnarray}
\label{solution}
a_{k\pi} = c_{k\pi}\frac{3}{8}[\eta^T + \frac{5}{3} \eta^{PS}],
\end{eqnarray}
with
\begin{eqnarray}
\label{c_pi}
c_{1\pi} = \frac{8}{3} h_{\pi} F_P \approx 132.4,\ \ \
c_{2\pi} = \frac{4}{3} h_{\pi}^2 \approx    170.3.
\end{eqnarray}
Here we accepted the conventional values of the current quark masses
$m_u=4.2$ MeV, $m_d = 7.4$ MeV.
The large ratio of the pion mass to the small current quark masses
provide an additional enhancement factor of the pion mechanism
as mentioned before the Eq.\ (\ref{2pi}).
Thus, we have obtained the approximate hadronic "image"
${\cal L}_{he}$ of the fundamental quark-lepton Lagrangian
${\cal L}_{qe}$ given in Eq.\ (\ref{ql}).

\section{Nuclear matrix elements of $R_p \hspace{-1em}/\;\:$ SUSY
induced $0\nu\beta\beta$-transition}
Starting from the Lagrangian ${\cal L}_{he}$ in Eq.\ (\ref{2pi})
it is straightforward to calculate
the ${0\nu\beta\beta}$-nuclear matrix element
\begin{eqnarray}
\label{DefL}
<(A, Z + 2), 2 e^-| S - 1|(A, Z)>\  =
<(A, Z + 2), 2 e^-| T exp[i \int d^4 x {\cal L}_{he}(x)] |(A, Z)>
\end{eqnarray}
The nuclear structure is involved via the initial (A,Z) and
the final (A, Z+2) nuclear states having the same atomic
weight A, but different electric charges Z and Z+2.
The nucleon-level diagrams, which correspond to the leading order
contributions to the amplitude in Eq. (\ref{DefL}), are given in Fig. 2.

The standard framework for the calculation of this nuclear matrix
element is the non-relativistic impulse approximation (NRIA) \cite{doi85}.

The final result for the half-life of $0\nu\beta\beta$-decay  in
$0^+\rightarrow 0^+$ channel with the two outgoing electrons in
the S-state, regarding all the three above-described
possibilities of hadronization, reads
\begin{eqnarray}
\label{susyhalflife}
&&\big[ T_{1/2}(0^+ \rightarrow 0^+) \big]^{-1} = \\ \nonumber
&& = G_{01} \left | \eta^T \cdot {\cal M}_{\tilde q}^{2N}
 +  (\eta^{PS} - \eta^T) \cdot {\cal M}_{\tilde f}^{2N} +
\frac{3}{8}(\eta^T + \frac{5}{3} \eta^{PS})
\left(\frac{4}{3} M^{1\pi} + M^{2\pi}\right)\right |^2\,.
\end{eqnarray}
Here $G_{01}$ is the standard phase space factor tabulated
for various nuclei in Ref. \cite{pa96}.
The nuclear matrix elements ${\cal M}_{\tilde q,\tilde f }^{2N}$
governing the sub-dominant two-nucleon mode were presented
in Ref.\cite{HKK96}. As was already mentioned its
contribution can be safely neglected.
The one- and the two-pion modes nuclear matrix elements
$M^{1\pi}$ and $M^{2\pi}$ we write down in the form
\begin{equation}
\label{susy2pime}
{\cal M}^{k\pi} = \left(\frac{m_A}{m_{_p}}\right)^2
\frac{m_{_p}}{ m_e}
      \alpha^{k\pi}\left(M_{GT}^{k\pi} + M_{T}^{k\pi} \right)
\end{equation}
Here, $m_A = 850$MeV is the mass scale of the nucleon form factor.

The structure coefficients $\alpha^{k\pi}$ in Eq.\  (\ref{susy2pime})
are related to the coefficients $c_{k\pi}$ introduced in
Eq.\ (\ref{c_pi}) and have the following explicit form
\begin{eqnarray}
\label{alpha-pi}
\alpha^{1\pi} &=& - 6 g_{_s} h_{\pi} \rho F_P \approx -4.4\cdot10^{-2},\\
 \alpha^{2\pi}&=& g_{_s}^2 h_{\pi}^2 \rho \approx 2.0\cdot 10^{-1},
\end{eqnarray}
where
\begin{eqnarray}
\label{rho1}
\rho = \frac{1}{36 f_A^2}\left(\frac{m_{\pi}}{m_{_p}}\right)^4
\left(\frac{m_{_p}}{m_A}\right)^2,
\end{eqnarray}
with $f_A \approx 1.261$ being the axial nucleon coupling.

The two types of the Gamow-Teller and tensor nuclear matrix elements
are given by the expressions
\begin{eqnarray}
\label{MGT}
M_{GT}^{k\pi} &=&
\langle 0^+_f|\sum_{i\neq j} \tau_i^+ \tau_j^+
{\bf \sigma}_i \cdot {\bf \sigma}_j
F_1^{(k)}(x_{\pi}) \frac{R}{r_{ij}}
|0^+_i\rangle\,,\quad\mbox{with}\quad k=1,2\\
\label{MT}
M_{T}^{k\pi}&=&
\langle 0^+_f|\sum_{i\neq j} \tau_i^+ \tau_j^+
\left[3({\bf \sigma}_i\cdot \hat{{\bf r}}_{ij})
       (\vec{\sigma_j}\cdot \hat{{\bf r}}_{ij})
      - {\bf \sigma}_i\cdot {\bf \sigma}_j\right]
      F_{2}^{(k)}(x_{\pi})\frac{R}{r_{ij}}
|0^+_i\rangle \,,
\end{eqnarray}
where
\begin{eqnarray}
\label{notation1}
x_\pi=m_\pi r_{ij},\ \ \  {\bf r}_{ij}={\bf r}_i-{\bf r}_j, \ \ \
r_{ij}=|{\bf r}_{ij}|,\ \ \
\hat{{\bf r}}_{ij}= \frac{{\bf r}_{ij}}{r_{ij}},
\end{eqnarray}
${\bf r}_i$ is a coordinate of the $i$th nucleon
and $R = r_0 A^{1/3}$ is the mean nuclear radius, with $r_0 = 1.1 fm$.

The structure functions $F_1^{(k)}(x_{\pi})$ and $F_2^{(k)}(x_{\pi})$
have their origin in the integration of the pi-meson propagators
and take the following form:
\begin{eqnarray}
F_1^{(1)}(x) &=&  e^{- x}, \ \ \ F_2^{(1)}(x) =
(3 + 3x + x^2)\frac{e^{- x}}{x^2},\\
F_1^{(2)}(x) &=& (x - 2) e^{- x}, \ \ \ F_2^{(2)}(x) = (x + 1) e^{- x}.
\label{potentials}
\end{eqnarray}

We calculate the nuclear matrix elements within
the proton-neutron renormalized
Quasiparticle Random Phase Approximation (pn-RQRPA) \cite{pn-RQRPA,TU}.
This nuclear structure method has been developed from
the proton-neutron QRPA (pn-QRPA)
approach, which has been frequently used in
the $0\nu\beta\beta$-decay calculations.
The pn-RQRPA is an extension of the pn-QRPA by incorporating the Pauli
exclusion principle for the fermion pairs.
The limitation of the conventional pn-QRPA is traced to the quasiboson
approximation (QBA), which violates the Pauli exclusion principle.
In the QBA one neglects the terms coming from the commutator of the
two bifermion operators by replacing the exact expression for this commutator
with its expectation value in the uncorrelated BCS ground state.
In this way the QBA implies the two-quasiparticle operator to be
a boson operator. The QBA leads to too strong ground state correlations
with increasing strength of the residual interaction in the
particle-particle channel what affects the calculated nuclear
matrix elements severely.
To overcome this problem the Pauli exclusion principle has to be
incorporated into the formalism \cite{pn-RQRPA}, \cite{TU} in order to
limit the number of quasiparticle pairs in the correlated ground state.
The commutator is not anymore boson like, but obtains corrections to
its bosonic behavior due to the fermionic constituents.
The pn-RQRPA goes beyond the QBA. The Pauli effect of
fermion pairs is included in the pn-RQRPA via the renormalized QBA (RQBA)
\cite{pn-RQRPA}, \cite{TU}, i.e.  by calculating the commutator of
two bifermion
operators in the correlated QRPA ground state. The RQBA was applied to
the $2\nu\beta\beta$-decay in Ref. \cite{pn-RQRPA} and to the
$0\nu\beta\beta$-decay for the first time in Ref. \cite{TU}.
Now it is widely recognized that the QBA  is a poor approximation and
that the pn-RQRPA offers the advantages over pn-QRPA. Let us stress that
there is no collapse of the pn-RQRPA solution for a physical value of
the nuclear force and that the nuclear matrix
elements have been found significantly less sensitive to the increasing
strength of the particle-particle interaction
in comparison with QRPA results \cite{si97}.
Thus, the pn-RQRPA provides significantly more reliable treatment of
the nuclear many-body problem for the description of the
$0\nu\beta\beta$ decay.

For numerical treatment of the $0\nu\beta\beta$-decay
matrix elements given in Eqs.\ (\ref{MGT}) and (\ref{MT})
within the pn-RQRPA we transform
them by using the second quantization formalism
to the form containing the  two-body matrix elements
in the relative coordinate. One obtains \cite{si97}:
\begin{eqnarray}
<{\cal O}_{ij}> =
\sum_{{p n p' n' } \atop {J^{\pi}
m_i m_f {\cal J}  }}
~(-)^{j_{n}+j_{p'}+J+{\cal J}}(2{\cal J}+1)
\left\{
\matrix{
j_p &j_n &J \cr
j_{n'}&j_{p'}&{\cal J}}
\right\}\times~~~~~~~~\nonumber \\
<p, p';{\cal J}|f(r_{ij})\tau_i^+ \tau_j^+ {\cal O}_{ij}
f(r_{ij})|n ,n';{\cal J}>\times ~~~~~~~~~~~~~
\nonumber \\
< 0_f^+ \parallel
\widetilde{[c^+_{p'}{\tilde{c}}_{n'}]_J} \parallel J^\pi m_f>
<J^\pi m_f|J^\pi m_i>
<J^\pi m_i \parallel [c^+_{p}{\tilde{c}}_{n}]_J \parallel
0^+_i >.
\end{eqnarray}
${\cal O}_{ij}$ represents the coordinate and spin dependent part of the
two body transition operators of the $0\nu\beta\beta$-decay
nuclear matrix elements in Eqs.\  (\ref{MGT}) and (\ref{MT}).
The short-range correlations
between the two interacting nucleons are taken into account by
the correlation function
\begin{equation}
\label{2nuccorr}
f(r)=1-e^{-\alpha r^2 }(1-b r^2) \quad \mbox{with} \quad
\alpha=1.1~ \mbox{fm}^2 \quad \mbox{and} \quad  b=0.68 ~\mbox{fm}^2.
\end{equation}
The one-body transition densities and the other details of the
nuclear structure model are given in \cite{si97,pn-RQRPA,TU}.

The calculated nuclear matrix elements for the $0\nu\beta\beta$-decay
of various isotopes within the pn-RQRPA are presented
in Table 1. The considered single-particle
model spaces both for protons and neutrons have been as follows:
i) For A=76, 82  the model space consists of the full
$2-4\hbar\omega$ major oscillator shells.
ii) For A=96, 100, 116 we added to the previous model space
$1f_{5/2}$, $1f_{7/2}$,
$0h_{9/2}$ and $0h_{11/2}$ levels.
iii) For A=128, 130, 136 the model space
comprises the full $2-5\hbar\omega$ major shells.
iv) For A=150 the model space extends over the full $2-5\hbar\omega$ shells
plus $0i_{11/2}$ and $0i_{13/2}$ levels.

The single particle
energies were  obtained by using  Coulomb corrected Woods Saxon
potential.  The interaction employed was the Brueckner G-matrix which is a
solution of the Bethe-Goldstone equation with the Bonn
one-boson exchange potential. Since the considered model space is
finite, we have renormalized pairing interactions by the strength
parameters $d_{pp}$ and $d_{nn}$ \cite{cheo93} to the empirical
gaps defined by Moeller and Nix \cite{mn92}.
The  particle-particle and particle-hole channels of the G-matrix
interaction of nuclear Hamiltonian $H$ have been renormalized with
parameters $g_{pp}$ and $g_{ph}$, respectively.
The  nuclear matrix elements listed
in the Table 1 have been obtained for the $g_{ph} = 0.80$ and
$g_{pp} = 1.0$

The following note is in order. According to our numerical analysis,
variations of the nuclear matrix elements presented in Table 1
do not exceed 20\%  within the physical region of
the nuclear structure parameter $g_{pp}$ ($0.8 \le g_{pp} \le 1.2$).

As seen from the Table 1 $M^{2\pi}$ is significantly larger
than $M^{1\pi}$. It is partially due to the mutual cancellation of the
$M_{GT}^{1\pi}$ and $M_{T}^{1\pi}$ in the construction
of $M^{1\pi}$ in Eq. (\ref{susy2pime})
(see Table 1)
and due to the suppression of the structure coefficient
$\alpha^{1\pi}$ in comparison with $\alpha^{2\pi}$.
Thus, we conclude that the two-pion mode
contribution to $0\nu\beta\beta$-decay (Fig. 2(c)) dominates both over
the one-pion (Fig. 2(b)) and the two-nucleon contributions
(Fig. 2(a)).
The dominance of the two-pion mode over the two-nucleon one
was previously proven in Ref. \cite{FKSS97}.

\section{Constraints on $R_p \hspace{-1em}/\;\:$ SUSY from
$0\nu\beta\beta$-experiments. Comparative analysis.}
Having all the quantities in the $0\nu\beta\beta$-decay half-life formula
(\ref{susyhalflife}) specified we are ready to extract the limits on
the $R_p \hspace{-1em}/\;\:$ parameters from non-observation of the
$0\nu\beta\beta$-decay.

 The experimental lower bound $T_{1/2}^{exp}$ for
the half-life of a certain isotope $Y$ provides
the following constraint on the effective $R_p \hspace{-1em}/\;\:$ SUSY
parameters
\begin{eqnarray}
\label{con}
\eta_{_{SUSY}}\equiv  \frac{3}{8}(\eta^T + \frac{5}{3} \eta^{PS})
 \leq \eta^{exp}_{_{SUSY}} =
        \frac{10^{-7}}{\zeta(Y)}
        \sqrt{\frac{10^{24} \mbox{years}} {T_{1/2}^{exp}}},
\end{eqnarray}
        Here we introduced the SUSY sensitivity $\zeta(Y)$ of
        a $0\nu\beta\beta$-decaying isotope $Y$
\begin{equation}
\label{sss}
\zeta(Y) = 10^{5}~ |\frac{4}{3} M^{1\pi} + M^{2\pi}|~ \sqrt{G_{01}}.
\end{equation}
        The quantity $\zeta(Y)$ is an
        intrinsic characteristic of
        an isotope $Y$ depending only on the nuclear matrix elements
        $M^{1\pi}$, $M^{2\pi}$ and on the phase space factor $G_{01}$.
        The large numerical values of the SUSY sensitivity $\zeta$ defined
        in (\ref{sss}) correspond to those isotopes within the group of
        $\beta\beta$-decaying nuclei which are the most promising
        candidates for searching SUSY in the $0\nu\beta\beta$-decay.

        The numerical values of  $\zeta(Y)$ calculated in the pn-RQRPA are
        presented in the Table 1 and displayed in Fig. 3 in the form of
        a histogram. It is seen that
        the most sensitive isotope is $^{150}$Nd, then follows
        $^{100}$Mo.

        It is understood that the SUSY sensitivity $\zeta$ cannot be
        the only criterion for selecting an isotope for
        the $0\nu\beta\beta$-experiment. Other microscopic and
        macroscopic properties of the isotope are also important
        for building a $0\nu\beta\beta$-detector.

        The current experimental situation in terms of the accessible
        half-life and the corresponding upper limit on the effective
        SUSY parameter $\eta_{_{SUSY}}$ is presented in Table 2.
        We conclude that the best upper limit on the
$R_p \hspace{-1em}/\;\:$ SUSY
        parameter $\eta_{_{SUSY}}$ has been established by
        the Heidelberg-Moscow experiment \cite{hdmo97}.
We denote this limit as $\eta_{_{SUSY}}^{exp}(H-M)$.
        For comparison in the bottom row of the Table 2 we show
        the lower half-life limits
$T^{exp}_{1/2}$($\eta^{H-M}_{_{SUSY}}$), which must be reached by
       $0\nu\beta\beta$-experiments  with the other nuclei
to reach this presently best constraint
$\eta_{_{SUSY}}\leq \eta_{_{SUSY}}^{exp}(H-M)$ on the
$R_p \hspace{-1em}/\;\:$ SUSY.
        The result of this comparison is illustrated in Fig. 3.

Using the values of $\eta_{_{SUSY}}^{exp}$ from the Table 2
one can easily calculate the corresponding
constraints on the $\lambda'_{111}$
parameter. There are two types of the constraints for each value of
$\eta_{_{SUSY}}^{exp}$  parameter
\begin{eqnarray}
\label{gn}
&&\lambda'_{111} \leq 1.8  \sqrt{\eta_{_{SUSY}}^{exp}}
\left(\frac{m_{\tilde q}}{100\mbox{GeV}}\right)^2
\left(\frac{m_{\tilde g}}{100\mbox{GeV}}\right)^{1/2},\\
\label{gn11}
&&\lambda'_{111} \leq 12.5 \sqrt{\eta_{_{SUSY}}^{exp}}
\left(\frac{m_{\tilde e}}{100\mbox{GeV}}\right)^2
\left(\frac{m_{\chi}}{100\mbox{GeV}}\right)^{1/2}.
\end{eqnarray}
These formulas are derived from Eqs.\ (\ref{etaq})-(\ref{eta_end}),
applying a widely used ansatz of the universal squark $\tilde q$ mass
$m_{\tilde q}$ at the weak scale $m_{\tilde u}\approx m_{\tilde d}\approx m_{\tilde q}$.
This approximation is well motivated by the constraints from
the flavor changing neutral currents.
Formula (\ref{gn11}) takes into account only the lightest
neutralino contribution with the mass $m_{\chi}$.
We also assume absence of spurious compensations between terms
of different nature such as the $\tilde g-q-\tilde q$ and
$\chi-e-\tilde e$. The running QCD coupling constant $\alpha_s(Q)$
has been taken at the scale $Q = 1$GeV with the normalization on the
world average value $\alpha_s(M_Z) = 0.120$ \cite{RPP}.
The second limit in Eq. \ (\ref{gn11}) is derived with
the additional assumptions that the lightest neutralino
is B-ino dominant and that
$m_{\tilde q} \geq m_{\tilde e}/2$. Both these assumptions
are phenomenologically reasonable, although they must not be always
correct.

The best constraint from the Heidelberg-Moscow experiment \cite{hdmo97}
is
\begin{eqnarray}
\label{limits}
&&\lambda'_{111} \leq  1.3 \cdot 10^{-4}\Big({m_{\tilde q}\over{100 GeV}}
\Big)^2
 \Big({m_{\tilde g}\over{100 GeV}} \Big)^{1/2},\\
&&\lambda'_{111} \leq 9.3 \cdot 10^{-4}
\Big({m_{\tilde e}\over{100 GeV}} \Big)^2
 \Big({m_{\chi}\over{100 GeV}} \Big)^{1/2}
\end{eqnarray}
These limits are very strong and, as it was already pointed out in
Ref. \cite{HKK96}-\cite{FKSS97-1}, lie beyond the reach of the near
future  accelerator experiments
(though, accelerator experiments are potentially sensitive
to the other couplings than $\lambda'_{111}$).

To constrain the size of $\lambda'_{111}$ itself one needs additional
assumptions on the masses of the SUSY-partners.
If the values of these masses would be around
their present experimental lower
limits $\sim 100$GeV \cite{RPP}, one could constrain
the coupling to
\begin{eqnarray}
\label{best}
\lambda'_{111}\leq 1.3 \cdot 10^{-4}.
\end{eqnarray}
A conservative bound can
be set by assuming all the SUSY-masses being at the "SUSY-naturalness"
bound of $\sim 1$TeV, leading to
\begin{eqnarray}
\label{natural}
\lambda'_{111}\leq 0.04.
\end{eqnarray}
This completes our analysis. The other details concerning
the experimental prospects for searching for $R_p \hspace{-1em}/\;\:$ SUSY
in $0\nu\beta\beta$-experiments
can be inferred directly from Table 2 and Fig. 3.

\section{Conclusion}
In summary, we have analyzed the general case of the
pion realization for the short-ranged $R_p \hspace{-1em}/\;\:$
SUSY mechanism of $0\nu\beta\beta$-decay taking into account
both the one-pion and two-pion modes. We have shown that
the two-pion mode $R_p \hspace{-1em}/\;\:$ SUSY contribution
to $0\nu\beta\beta$-decay dominates over the one-pion mode contribution.
Previously \cite{FKSS97} we had proven that the two-pion
mode dominates over the conventional two-nucleon one.
We also pointed out that non-observation of $0\nu\beta\beta$-decay
casts severe limitations on the $R_p \hspace{-1em}/\;\:$ SUSY
extensions of the standard model of electroweak interaction.
Although a complicated nuclear many-body problem needs to be
solved the limits
are so stringent that they overcome the uncertainties in
the nuclear and hadronic matrix elements, leading to limits
that are much stronger than those  from accelerator
and the other non-accelerator experiments.
We gave the list of nuclear matrix elements for all the experimentally
interesting isotopes and  presented the so called
SUSY sensitivities of these isotopes.
These characteristic might be helpful
for planning future searches for SUSY in $0\nu\beta\beta$-decay.
On this basis we compared the present status of the various
$0\nu\beta\beta$-experiments and their abilities to detect the SUSY signal.

\centerline{\bf ACKNOWLEDGMENTS}
\bigskip
We are grateful to V.A. Bednyakov for helpful discussions.
S.K. would like to thank the "Deutsche Forschungsgemeinschaft"
for financial support by grant Fa 67/21-1. The research described in this
publication was made possible in part by EU support under contract
CT94-0603, by Grant Agency of Czech. Rep. contract No. 202/98/1216
and by grant GNTP 315NUCLON from the Russian ministry of science.

\newpage

\begin{table}[t]
\caption{Nuclear matrix elements for the pion-exchange
R-parity violating SUSY mode of $0\nu\beta\beta$-decay for the
experimentally most interesting isotopes calculated within
the renormalized
pn-QRPA. $G_{01}$ is the integrated kinematical factors for
$0^+ \rightarrow 0^+$ transition [18].
$\zeta (Y)$ denotes according to Eq. (\ref{sss}) the sensitivity of
a given nucleus $Y$ to the SUSY signal.}
\begin{tabular}{lllllllll}
Nucleus  & ${\cal M}^{1\pi }_{GT}$& ${\cal M}^{1\pi }_{T}$ &
         ${\cal M}^{2\pi }_{GT}$ &${\cal M}^{2\pi }_{T}$ &
         ${\cal M}^{1\pi}$ &${\cal M}^{2\pi}$ &
         $G_{01}$    & $\zeta (Y) $\\
&&&&&&&$\times 10^{15} y$ &\\
\hline
$^{76}Ge$  &1.30 &-1.02 &-1.34 &-0.65 &-18.2 &-601 &7.93 &5.5 \\
$^{82}Se$  &1.23 &-0.87 &-1.26 &-0.57 &-23.9 &-551 &35.2 &10.8\\
$^{96}Zr$  &0.77 &-1.11 &-0.85 &-0.67 &22.1  &-458 &73.6 &11.8\\
$^{100}Mo$ &1.43 &-1.73 &-1.52 &-1.05 &19.4  &-776 &57.3 &18.1\\
$^{116}Cd$ &0.92 &-0.78 &-0.94 &-0.47 &-9.3  &-423 &62.3 &10.8\\
$^{128}Te$ &1.25 &-1.57 &-1.40 &-0.99 &21.4  &-720 &2.21 &3.3 \\
$^{130}Te$ &1.10 &-1.48 &-1.26 &-0.93 &25.1  &-660 &55.4 &14.9\\
$^{136}Xe$ &0.61 &-0.84 &-0.74 &-0.54 &15.5  &-387 &59.1 &9.0 \\
$^{150}Nd$ &1.85 &-2.70 &-2.07 &-1.68 &56.4  &-1129&269. &55.6\\
\end{tabular}
\label{table.1}
\end{table}

\begin{table}[t]
\caption{The present state of the
$R_p \hspace{-1em}/\;\:$ SUSY searches in
$\beta\beta$-decay experiments.
$T^{exp}_{1/2}$(present) is
the best presently available lower limit on
the half-life of the $0\nu\beta\beta$-decay for a given isotope.
$\eta_{_{SUSY}}^{exp}$ is the corresponding upper limit
on the $R_p \hspace{-1em}/\;\:$ SUSY parameter.
$T^{exp}_{1/2}$($\eta^{H-M}_{_{SUSY}}$) is the  calculated half-life
of $0\nu\beta\beta$-decay
assuming  $\eta_{_{SUSY}} =\eta^{H-M}_{_{SUSY}}$
with $\eta^{H-M}_{_{SUSY}}$ being
the best limit deduced from the Heidelberg-Moscow
$^{76}Ge$ experiment [1].
}
\begin{tabular}{llllll}
Nucleus  & $^{76}Ge$ & $^{82}Se$ & $^{96}Zr$ & $^{100}Mo$ & $^{116}Cd$ \\
Ref. & \cite{hdmo97} & \cite{ell92} & \cite{kaw93} & \cite{eji96} &
       \cite{dane95} \\ \hline
 $T^{exp}_{1/2}$(present) &
 $ 1.1\times 10^{25}$    & $  2.7\times 10^{22}$ &
 $ 3.9\times 10^{19}$    & $  5.2\times 10^{22}$ &  $ 2.9\times 10^{22}$ \\
 $\eta^{exp}_{_{SUSY}}$      & $ 5.5\times 10^{-9} $ & $ 5.6\times 10^{-8} $ &
 $ 1.4\times 10^{-6} $ & $ 2.4\times 10^{-8} $ & $ 5.4\times 10^{-8} $
\\[1mm]
 $T^{exp}_{1/2}$($\eta^{H-M}_{_{SUSY}}$) &
  $ 1.1\times 10^{25}$    & $ 2.9\times 10^{24}$ &
  $ 2.4\times 10^{24}$    & $ 1.0\times 10^{24}$ &    $2.9\times 10^{24}$ \\
 & & & & & \\
\hline
Nucleus & $^{128}Te$ & $^{130}Te$ & $^{136}Xe$ & $^{150}Nd$ & \\
      & \cite{bern92} & \cite{ale94} & \cite{bus96} & \cite{sil97} & \\
\hline
 $T^{exp}_{1/2}$(present) &
 $  7.7\times 10^{24}$   &  $  8.2\times 10^{21}$  &
 $  4.2\times 10^{23}$   &  $  1.2\times 10^{21}$  &  \\
 $\eta^{exp}_{_{SUSY}}$      & $ 1.1\times 10^{-8} $ & $ 7.4\times 10^{-8} $ &
 $ 1.7\times 10^{-8} $ & $ 5.2\times 10^{-8} $ &  \\[1mm]
 $T^{exp}_{1/2}$($\eta^{H-M}_{_{SUSY}}$) &
    $ 3.1\times 10^{25}$ &   $ 1.5\times 10^{24}$  &
    $ 4.1\times 10^{24}$ &   $ 1.1\times 10^{23}$  & \\
\end{tabular}
\label{table.2}
\end{table}

\bigskip
{\large\bf Figure Captions}

\bigskip
\noindent
{\bf Fig. 1:}
An example of the supersymmetric contribution
to ${0\nu\beta\beta}$-decay with the gluino $\tilde g$ and
two squarks $\tilde u$ in the intermediate state.

\bigskip
\noindent
{\bf Fig. 2:}
The hadronic-level diagrams for the short-ranged SUSY mechanism of
$0\nu\beta\beta$-decay.
(a) the conventional two-nucleon mode, (b) the one-pion exchange mode,
(c) the two-pion exchange mode.
\bigskip

\noindent
{\bf Fig. 3:}
The SUSY sensitivity $\zeta (Y)$ for the experimentally interesting
nuclei (on the left).
This histogram displays the corresponding numerical values in the Table 1.
The histogram on the right illustrates the Table 2.
The best presently available lower limits on the $0\nu\beta\beta$-decay
half-life $T^{exp}_{1/2}$ are denoted by the black bars.
The open bars indicate the half-life limits
$T^{exp}_{1/2}(\eta^{H-M}_{_{SUSY}})$ to be reached
by a given experiment to reach the presently best
limit on the $R_p \hspace{-1em}/\;\:$ SUSY parameter
$\eta^{H-M}_{_{SUSY}}$ established by
the $^{76}$Ge experiment \cite{hdmo97}.

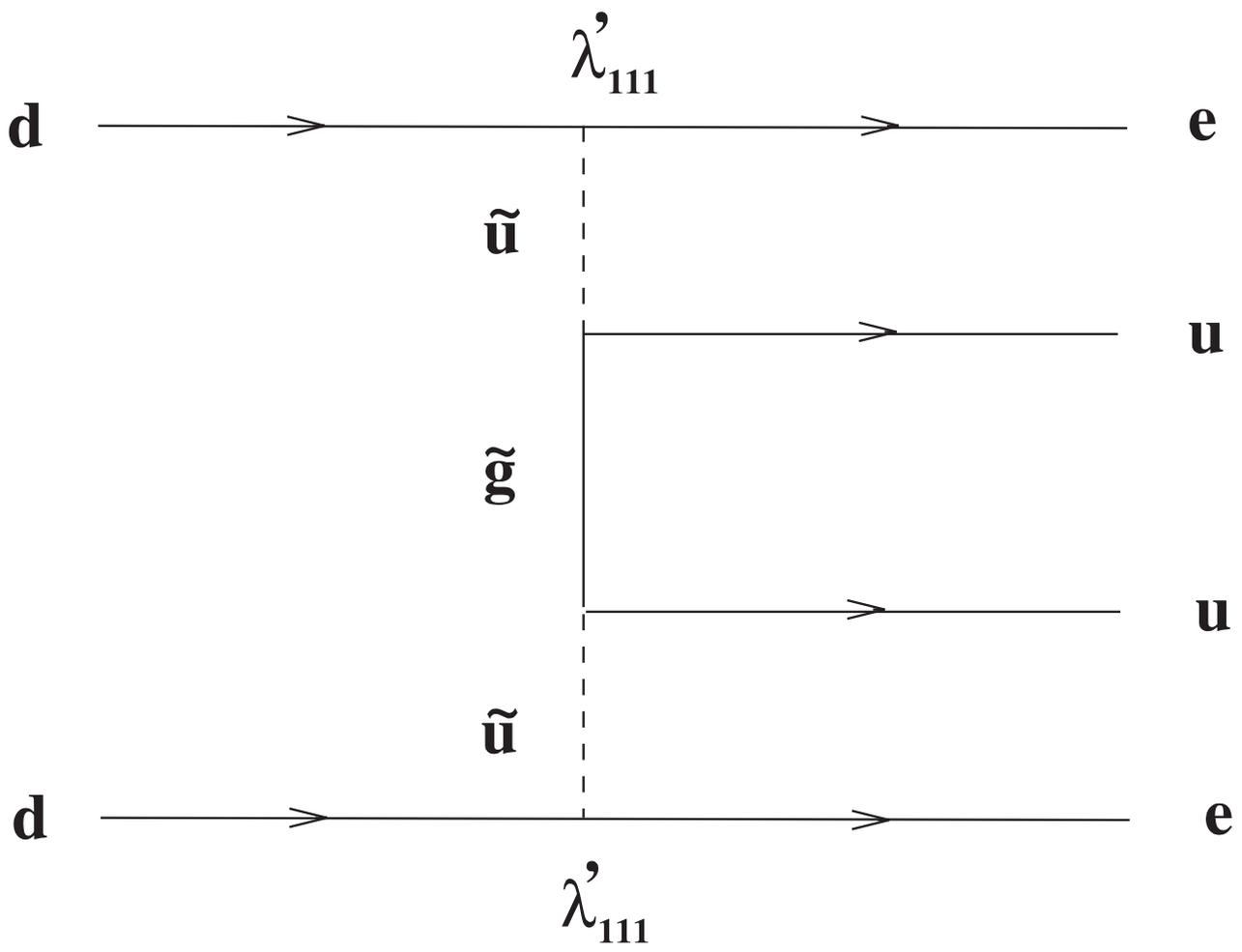
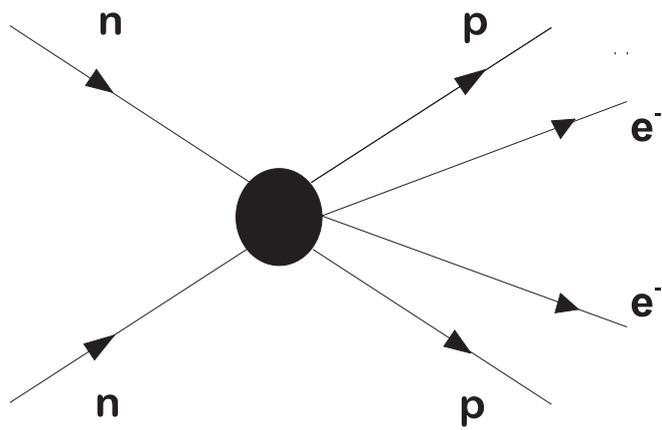
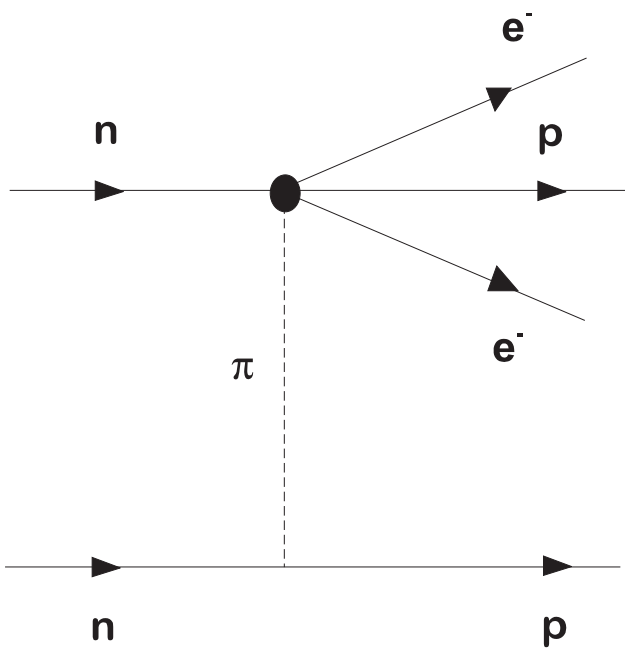
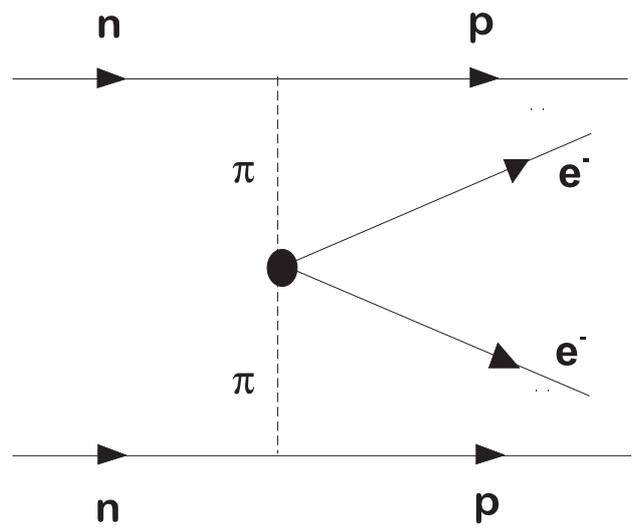
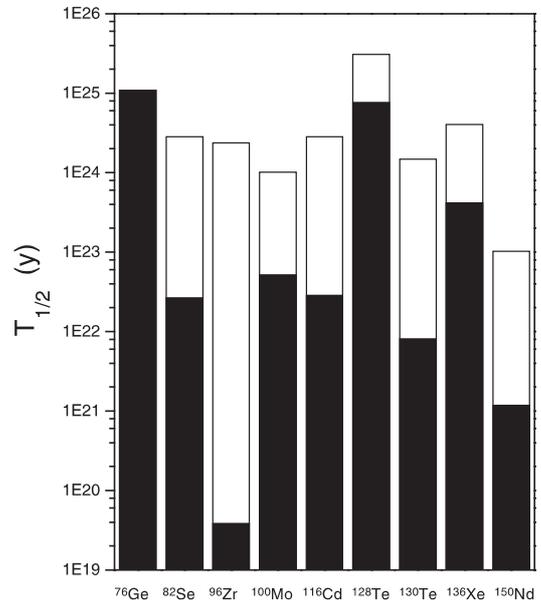
\end{document}